\documentstyle[12pt,aps]{revtex}
\input{psfig.sty}

\begin{document}
\title{Lattice-dynamics of a disordered solid-solid interface}
\author{G. Fagas, A.G. Kozorezov, C.J. Lambert, J.K. Wigmore}
\address{School of Physics and Chemistry, Lancaster University, LA1 4YB, UK}
\author{A. Peacock, A. Poelaert, R. den Hartog}
\address{Astrophysics Division, European Space Agency-ESTEC, Noordwijk,
The Netherlands}
\maketitle

\begin{abstract}
Generic properties of elastic phonon transport at a disordered interface
are studied. The results show that phonon transmittance is a strong function
of frequency and the disorder correlation length. At frequencies lower than the
van Hove singularity the transmittance at a given frequency increases
as the correlation length decreases. At low frequencies, this is reflected
by different power-laws for phonon conductance across correlated
and uncorrelated disordered interfaces which are in approximate agreement
with perturbation theory of an elastic continuum.
These results can be understood in terms of simple mosaic and two-colour
models of the interface.
\end{abstract}

\section{Introduction}
The transfer of phonons across the interface between two solids
is crucial to the operation of a range of novel devices from
semiconductor nanostructures,
quantum-cascade lasers, vertical structures, superlattices,
quantum dot arrays to the cryogenic phonon-mediated detectors of
elementary particles. In such structures non-radiative transitions and
hot-electron thermalization gives rise to fluxes of highly
energetic phonons spanning the whole of the Brillouin zone.

In early studies \cite{Little}, phonons were regarded as bulk elastic
continuum waves so that transmission and reflection of energy
at the interface between two materials were a consequence
of acoustic mismatch. Such a macroscopic continuum theory
describes generic low-frequency properties but does not account for
important features such as phonon dispersion, the Debye cutoff and weak bonding
at the interface, for which a lattice-dynamical model is required \cite{Maris}.
Most early investigations assume that the solid-solid interface
possesses in-plane translational invariance, whereas, it is known that
disorder-induced scattering can play a significant role \cite{WW,Pohl1}.
Within a lowest order perturbation theory  \cite{Alex}, it is
predicted that disorder can induce a strong frequency dependence in phonon
transport coefficients with a power-law crossover determined by interface
conditions. The aim of this paper is to go beyond this elastic
continuum perturbative approach, by formulating an exact lattice-description
valid for terahertz phonon transport, which includes features such as
phonon dispersion and singularities in the phonon density of states.

To formulate the problem of phonon transmission (reflection)
we consider two identical
semi-infinite leads of uniform cross-section attached to a 'scattering region'.
The two semi-infinite perfect crystals are treated as waveguides
for incident and scattered phonons whose dynamics is described by
the lattice equations of motion. The geometry analyzed is shown in Fig.1,
where the scattering region
consists of a single disordered atomic plane.
Disorder at the planar interface is introduced microscopically as correlated
or uncorrelated random variation of the on-site atomic masses
and we assume that there are no significant anharmonic effects.
To calculate transport properties of
such a system, we use methods specifically developed to investigate
electronic transport through
phase-coherent structures  \cite{Landauer,Lambert2,Stefano}.
Transport coefficients are calculated as a function of 
phonon frequency within the Landauer-Buttiker formalism \cite{Landauer}
by employing an exact recursive Green function technique \cite{Stefano}.

\section{Details of the calculation}
In what follows we present the results for two lattice dynamical models:
an fcc lattice with central-force nearest-neighbour interactions and a
scalar square lattice. For the former the atomic masses
$M_{lmn}$ are located at equilibrium positions ${\bf r}_{lmn} = (la,ma,na)$,
where $l,m,n$ sum to an even number and $a {\sqrt 2}$ is the
nearest-neighbour spacing, whereas for the square lattice $a$
is the lattice constant and $n=0$. In the latter case we do not distinguish
between phonon polarizations and hence, are not concerned about
mode conversion due to scattering.

In the harmonic approximation the linearized equations of motion
of an atom located at ${\bf r}_{lmn}$ is

\begin{equation}
M_{lmn}\ddot{u}_{lmn}^{\alpha }=-\sum_{\beta,l^\prime m^\prime n^\prime}
K_{lmn,l^\prime m^\prime n^\prime}^{\alpha \beta }
u_{l^\prime m^\prime n^\prime}^{\beta },
\label{motion}
\end{equation}

where $u_{lmn}^{\alpha }$ is the atomic displacement vector,

\begin{eqnarray}
K_{lmn,l^\prime m^\prime n^\prime}^{\alpha \beta}
& = &\kappa ( s_{lmn,l^\prime m^\prime n^\prime}^{\alpha}
s_{lmn,l^\prime m^\prime n^\prime}^{\beta}-
4\delta_{lmn, l^\prime m^\prime n^\prime}
\delta^{\alpha \beta} ), \nonumber \\
s_{lmn,l^\prime m^\prime n^\prime}^{\alpha } & = & \frac
{ {\bf r}_{l^\prime m^\prime n^\prime}^{\alpha}-{\bf r}_{lmn}^{\alpha} }
{ \left| {\bf r}_{l^\prime m^\prime n^\prime}-{\bf r}_{lmn} \right| },
\label{force}
\end{eqnarray}

and ${\bf r}_{l^\prime m^\prime n^\prime}-{\bf r}_{lmn}$
is taken from the first coordination sphere. The indices $\alpha,\beta$ label 
the $x,y,z$ components of displacements for the fcc model
and should be dropped for the scalar model.

All atoms in the semi-infinite leads are of unit mass, whereas
the random masses at the interface have mean value unity
and standard deviation $\sigma$. For simplicity we allow the masses
to fluctuate only along a single direction, which is chosen to be
the [101] and [100] cubic crystallographic directions
for the fcc and square lattice respectively.
Henceforth, we refer to this as the 'disorder-line' and denote masses
at different positions along the line by $M_j$.
Thus, even for a three dimensional lattice,
due to the residual translational invariance, scattering is
confined in the plane fixed by the disorder-line and the angle
with respect to the interface determined by the conserved component
of the incident phonon momentum.
From this point of view our scattering problem is essentialy
two dimensional.

An ensemble of $N$ positive masses, where $N$ is the number of atoms along the 
direction of the disorder-line in the lead cross-section, is generated as
follows: First we introduce the correlated random numbers

\begin{equation}
\chi_{j} = \sum_{k} (a_{k}\cos(kj) + b_{k}\sin(kj)),
\label{random1}
\end{equation}

where $a_{k}$ and $b_{k}$ are Gaussian random numbers with zero mean value
and $\langle a_{k}a_{k^\prime}\rangle = \langle b_{k}b_{k^\prime}\rangle =
\delta_{kk^\prime}e^{-\frac{k^2 \xi^2}{4}}$. $k$ is taken commensurate
with the phonon wavenumbers determined by the boundary conditions in the
transverse direction and
$\xi$ denotes the correlation length. Random masses $M_j$
along the disorder-line with mean value unity
and standard deviation $\sigma$ are obtained from $\chi_{j}$ 
by the linear transformation

\begin{equation}
M_j =  1 + \sigma \cdot \frac {\chi_j - \langle \chi_j \rangle}
{\sqrt {\langle \chi_j^2 \rangle - {\langle \chi_j \rangle}^2 }}.
\label{random2}
\end{equation}

In order to isolate the effect of correlations, we compare phonon
transmission (reflection) coefficients for such correlated interfacial
disorder, with those of an uncorrelated
disorder configuration obtained by randomly 'shuffling' the above set.

\section{S-matrix and phonon interfacial transmittances}
In the absence of inelastic scattering,
phonon transport through an arbitrary scattering
region can be described by the scattering matrix
$S$ \cite{Landauer}, which depends on the conserved phonon frequency and
yields probabilities for all
scattering transitions within a system described by a dynamical matrix
$D_{lmn,l^\prime m^\prime n^\prime}^{\alpha \beta}=
M_{lmn}^{-1/2}K_{lmn,l^\prime m^\prime n^\prime}^{\alpha \beta}
M_{l^\prime m^\prime n^\prime}^{-1/2}$. If the scattering region is connected
to external reservoirs by phonon waveguides with
open channels labelled by quantum numbers $\nu$, then
$S$-matrix elements $S_{\nu\nu^\prime}(\omega,D)$ are defined such
that $|S_{\nu\nu^\prime}(\omega,D)|^2$ is the outgoing flux of phonons
along channel $\nu$, arising from a unit flux incident along channel
$\nu^\prime$. The scattering matrix defined this way is
unitary, i.e., $SS^\dagger=1$, and since
$S_{\nu\nu^\prime}(\omega,D)=S_{\nu^\prime\nu}(\omega,D^\dagger)$,
where $D$ is real, $S(\omega,D)$ is symmetric.
If $\nu,\nu^\prime$ belong to the left (right) lead then the
corresponding $S$-matrix elements are left (right) reflection amplitudes
which we denote $r_{\nu\nu^\prime}=S_{\nu\nu^\prime}(r_{\nu\nu^\prime}^\prime=
S_{\nu\nu^\prime})$, whereas, if $\nu,\nu^\prime$ belong to the right (left)
and left (right) lead, the $S$-matrix elements are
left (right) transmission amplitudes denoted
$t_{\nu\nu^\prime}=S_{\nu\nu^\prime}(t_{\nu\nu^\prime}^\prime=
S_{\nu\nu^\prime})$. It is evident that for an $N^\prime$-mode waveguide
three distinct transmittances can be defined,

\begin{equation}
T_{\nu\nu^\prime}=|t_{\nu\nu^\prime}|^2,\;\;
T_{\nu^\prime}=\sum_{\nu=1}^{N^\prime} |t_{\nu\nu^\prime}|^2,\;\;
T=\sum_{\nu,\nu^\prime=1}^{N^\prime} |t_{\nu\nu^\prime}|^2,
\label{trans1}
\end{equation}

To isolate the frequency dependence of the overall phonon
transmittance it is natural to normalize $T$ by the density of states.
In the presence of finite-width leads this means averaging $T$
over all open scattering channels $N^\prime$ and is equivalent to the
angle-averaged transmission coefficient,

\begin{equation}
\langle T(\omega,D) \rangle =
T(\omega,D)/{N^\prime},
\label{trans2}
\end{equation}

which is related to the corresponding reflection coefficient by

\begin{equation}
\langle R(\omega,D) \rangle=1-\langle T(\omega,D) \rangle.
\label{refl1}
\end{equation}

In what follows the $S$-matrix is obtained by solving
Dyson's equation for the Green function at the
surfaces of the two leads \cite{Lambert2,Stefano}
(see Fig.1 for definition of L,R), namely

\begin{equation}
G(\omega^2)=\left(\begin{array}{cc} G(L,L;\omega^2) & G(L,R;\omega^2)\\
G(R,L;\omega^2) & G(R,R;\omega^2)\end{array}\right),
\label{Green}
\end{equation}

\begin{equation}
G(\omega^2)=(g(\omega^2)^{-1}-D_{eff}(\omega^2))^{-1},
\label{Dyson}
\end{equation}

where $g(\omega^2)$ is the surface Green function of the
semi-infinite leads

\begin{equation}
g(\omega^2)=\left(\begin{array}{cc} g(L,L;\omega^2) & 0\\
0 & g(R,R;\omega^2)\end{array}\right),
\label{leads}
\end{equation}

and $D_{eff}(\omega^2)$ an effective dynamical matrix

\begin{equation}
D_{eff}(\omega^2)=
\left(\begin{array}{cc} D_{eff}(L,L;\omega^2) & D_{eff}(L,R;\omega^2)\\
D_{eff}(R,L;\omega^2) & D_{eff}(R,R;\omega^2)\end{array}\right).
\label{decim}
\end{equation}

The latter describes the effective (renormalized) coupling between the
surfaces of the two leads, obtained by projecting out the
internal degrees of freedom of the scatterer using the recursive Green function
technique. The Green
function $g(\omega^2)$, the spectrum and structure of the allowed wavequide
modes at fixed phonon density of states, as well as $S(\omega,D)$ are
calculated using a general algorithm developed in Ref. \cite{Stefano}.

\section{Phonon transmittance at low frequencies}
We present results for the average transmission coefficient
$\langle T(\omega) \rangle _c$
where the subscript c denotes a configurational averaging.
In Fig.2(a),(b) the frequency dependence of the
overall transmission coefficient $\langle T(\omega) \rangle _c$ averaged over
10 realizations of the interfacial mass-disorder
for the fcc lattice and over 500 for the scalar square lattice is shown.
The width of the disorder-line is 29 and 100 nearest-neighbour spacings
respectively, which corresponds to $3 \times 29$ and $1 \times 100$
degrees of freedom.

For the fcc model periodic boundary conditions in the lateral direction
(disorder-line) were used, whereas for the scalar model we adopted
fixed-end boundary conditions.
The boundary conditions set the precise frequencies at which
propagating waveguide modes become available in the external leads,
but apart from features which scale inversely with the number of such
modes do not affect the shape of the transmittance curves.
Figure 2 shows that for both fcc and scalar lattices, phonon
transmittances (reflectances) follow the same universal behaviour.
At sufficiently low frequencies, the phonon transmittance for correlated
disorder is lower than for uncorrelated interfacial disorder with
the same mean mass and standard deviation $\sigma$.

At low frequencies the results of lattice-dynamical calculations in
Fig.2(a),(b) are in good agreement with
perturbation theory \cite{Alex}. To model the disorder in
\cite{Alex} a thin layer of dissimilar material
sandwiched between irregular boundaries was introduced. The density
and elastic constants of this layer were assumed to differ from those
of the bulk on either side of the layer. To make a comparison with the
perturbative analysis we consider the case of
no fluctuation in the elastic constants and a line of random masses
at the location of the idealized interfacial plane $z=0$, i.e.,
$\Delta \rho ({\bf x},z) = \Delta \rho \zeta ({\bf x}) \delta (z) =
\Delta \rho ({\bf x}) \delta (z)$, with
$\zeta ({\bf x})$ a random function of lateral coordinate with zero mean value.
Apart from lattice discreteness which plays no significant role in
the long-wavelength limit there is no difference between our model of
site-to-site mass disorder and the model of interfacial roughness of 
Ref. \cite{Alex}. The following relation is useful when comparing the results

\begin{eqnarray}
\langle { \left(\frac{\Delta M_j}{M}\right)}^2 \rangle = {\sigma}^2 =
\left(\frac{\Delta \rho}{\rho}\right)^2 \; \frac{\langle {\zeta}^2 \rangle}{a^2}.
\label{map}
\end{eqnarray}

For such a two dimensional model the perturbation theory \cite{Alex} with
uncorrelated disorder $\xi=a$ and $k \xi \ll 1$, where k is the
phonon wavevector, yields
\begin{eqnarray}
\langle T(\omega) \rangle _c^{unc} = 1 - \pi^{\frac{5}{2}} \; \sigma^2
\; \left( \frac{\omega}{\omega_{max}} \right)^3,
\label{per1}
\end{eqnarray}

while for correlated disorder at sufficiently high frequencies
so that $k \xi \gg 1$ the perturbative result is

\begin{eqnarray}
\langle T(\omega) \rangle _c^{cor} = 1 - \pi^2 \; \sigma^2
\; \left( \frac{\omega}{\omega_{max}} \right)^2.
\label{per2}
\end{eqnarray}

More generally for all $\omega$, Fig.2(c) shows the
perturbative results for $\xi \rightarrow 0$ and $\xi \rightarrow \infty$.
Equations (\ref{per1}) and (\ref{per2}) are in
good agreement with the exact lattice-dynamical calculation.
They both show lower transmittance
for correlated disorder over a wide range of frequencies until the faster
power dependence of phonon transmittance for uncorrelated disorder
causes the two curves to cross at a frequency $\omega^\star$.
According to the perturbative calculations $\omega^\star$ should be
close to half the maximum phonon frequency $\omega_{max}$
independent of disorder characteristics, correlation length
and strength. Despite the similarities, this is not evident in the
lattice-dynamical calculations. Indeed, the crossing point
is quite close to a singularity in the phonon density of states
and thus using an elastic continuum model at frequencies of order $\omega^\star$
is an oversimplification.

Another simplifying assumption of the
theory in Ref.\cite{Alex} is the neglect of multiple scattering. This gives
the power law
$\langle R(\omega) \rangle _c^{unc}\sim\omega^{d+1}$ (Rayleigh scattering)
with subsequent decrease by one (two) power(s) for correlated disorder in
2D (3D). A log-log plot of phonon reflectances
showed that both lattice dispersion and multiple
phonon scattering effects result in a change of the power exponents
to a slightly different value. However, the drop by approximately
one power for our 2D scattering problem in the case of
correlated disorder prevails in all cases. The effect is attributed
to the restrictions in the phase volume available for the scattered
states, due to the correlation-induced finite width of the disordered
spectral distribution.

\section{Spectral properties of phonon transmittance}
Phonon transmittances over the whole frequency spectrum for the scalar model
for two different values of disorder strength $\sigma$ are shown in
Fig.3. In addition to the low frequency behaviour discussed
in Sec. IV, the transmittance shows a cusp at a frequency
corresponding to a van Hove singularity in the phonon density of states
and rapidly drops to zero at the maximum frequency
of the phonon spectrum $\omega_{max}$.
At the van Hove singularity the difference between correlated and
uncorrelated disorder realizations is much less pronounced.
All these features are also seen for the fcc model but
the exact behaviour at van Hove singularities is less clear due to
the finite-size effects described in the previous section.
To interpret the results we concentrate on the scalar
model.

The general expression for the transmission amplitude from incident
channel $\nu^\prime$ to channel $\nu$ for systems with nearest-neighbour
interactions has been derived in \cite{Stefano}. The
explicit expression for the scalar model with periodic boundary conditions is

\begin{eqnarray}
t_{\nu\nu^\prime}=\frac{1} {N} (\sum_{m,m_1,m^\prime=1}^N
e^{-ik_\nu^yma}G_{LR}(m,m_1)W(m_1,m^\prime)
e^{ik_{\nu^\prime}^ym^\prime a)})
\sqrt{ \frac{v_\nu} {v_{\nu^\prime}} }e^{-ik_\nu^xa}.
\label{amplit}
\end{eqnarray}

Here ${v_\nu}$ is the component of the group velocity in {\it x}-direction
along the lead axis of the $\nu$-th channel mode, m labels sites at the
interface (along lateral direction), and $G_{LR}$ has been defined in
Eq.(\ref{Green}). The matrix W is given by \cite{Stefano}

\begin{eqnarray}
W(m_1,m^\prime)=
\frac{\omega_{max}^2} {8N} \sum_{\mu=1}^N e^{ik_\mu^y(m_1-m^\prime )a}
(e^{ik_\mu^xa}-e^{-ik_\mu^xa}).
\label{W}
\end{eqnarray}

In formulae (\ref{amplit},\ref{W}), $k_\nu^y = \frac{2\nu\pi}{Na}$, as a result
of transverse quantization with periodic boundary conditions, while
$k_\nu^x$ denotes the wavenumber along the lead axis for phonons
of fixed frequency $\omega$ of state $\nu$ ($\nu$-th channel).

Using the complete orthonormalized set of functions
$\{\frac{1} {\sqrt{N}}e^{ik_\mu^yja}\}$ describing the profile of modes
in the transverse direction, $G_{LR}(m,m_1)$ can be written

\begin{eqnarray}
G_{LR}(m,m_1)=\frac{1} {N} \sum_{\mu,\mu^\prime=1}^N e^{ik_{\mu}^yma}
G_{LR}(\mu,\mu^\prime)e^{-ik_{\mu^\prime}^ym_1a}.
\label{Green1}
\end{eqnarray}

By substitution of Eq.(\ref{W}),(\ref{Green1}) to (\ref{amplit})
one obtains

\begin{eqnarray}
t_{\nu\nu^\prime}=\frac{\omega_{max}^2} {4}
G_{LR}(\nu,\nu^\prime)\;i\; sin(k_{\nu^\prime}^xa)
\sqrt{ \frac{v_\nu} {v_{\nu^\prime}} }e^{-ik_\nu^xa}.
\label{amplit1}
\end{eqnarray}

Thus, to calculate the channel-to-channel transmittance
${\langle T_{\nu\nu^\prime} \rangle}_c
={\langle |t_{\nu\nu^\prime}|^2 \rangle}_c$ or generally the
overall transmittance ${\langle T \rangle}_c$
one needs to consider the non-trivial statistically
averaged product of two Green functions,
${\langle G_{LR}(\nu,\nu^\prime)G_{LR}^\ast(\nu,\nu^\prime)\rangle}_c$
or ${\langle G_{LR}(m,m^\prime)G_{LR}^\ast(s,s^\prime)\rangle}_c$.

In what follows, we shall demonstrate that generic features can be captured
by considering the case of a regular array of atoms with mass $M^\prime$
located at the interface, thereby forming a linear(planar) defect
in a two(three) dimensional system.

In this case, the full Green function $G_{LR}(\nu,\nu^\prime)$
is of the form

\begin{eqnarray}
G_{LR}(\nu,\nu^\prime) =
\frac {g_o(\nu, \nu^\prime)e^{ik_{\nu^\prime}^xa}}
{1-\sigma^\prime\omega^2 g_o(\nu, \nu^\prime)},
\label{Green2}
\end{eqnarray}

where $g_o(\nu, \nu^\prime) = -4i\delta_{\nu, \nu^\prime}
/(\omega_{max}^2 sin(k_{\nu^\prime}^x))$
is the Green function of the infinite ideal lead at coinciding arguments,
and $\sigma^\prime=\frac{M^\prime-M}{M}$ is the relative
mass difference at the location of the defects.
Substituting this result into Eq.(\ref{trans2}) and (\ref{amplit1})
yields for the transmittance

\begin{eqnarray}
\langle T \rangle =\frac{1}{N^\prime} \sum_{\nu(open)}
\frac {sin^2(k_{\nu}^xa)} {sin^2(k_{\nu}^xa) + {\sigma^\prime}^2
(\frac {2\omega}{\omega_{max}})^4}
\label{total}
\end{eqnarray}

Eq.(\ref{total}) exhibits the following features:
(i)$\langle T \rangle$ equals unity for an ideal
interface with $\sigma^\prime=0$.
(ii)At low frequencies both $\omega, k_\nu^x \rightarrow 0$,
$sin(k_\nu^xa)\sim\omega$ and therefore
$\langle T \rangle \rightarrow 1-\alpha\omega^2$, where
$\alpha$ is numerical coefficient, in full agreement with the
results of perturbation theory \cite{Alex} and formula (\ref{per2}).
(iii)At the end of the phonon spectrum $\langle T \rangle \rightarrow 0$
since propagating channels are closing with $sin(k_\nu^xa)$
approaching zero at the Brillouin zone boundaries. This effect is
analogous to the total reflection by a single mass defect in one dimension.
(iv)At the van Hove singularity all channels are open and Eq.(\ref{total})
reduces to

\begin{eqnarray}
\langle T \rangle=\frac{1}{N} \sum_{\nu=1}^N
\frac {sin^2( \frac{2\nu\pi}{N})} {sin^2( \frac{2\nu\pi}{N})
+ 4{\sigma^\prime}^2},
\label{Hove}
\end{eqnarray}

which suggests that for $\sigma^\prime<1$ the dependence on disorder
strength is much weaker than at low frequencies. Also, $sin(k_\nu^ya)$ is small
for a large number of contributing channels thereby yielding
the characteristic cusp.
(v)We also note that Eq.(\ref{total}) has been derived for the [100] orientation
of the interface but since the phonon group velocity along the
lead axis is proportional to $sin(k_\nu^xa)$
the transmittance is in general anisotropic.

In Fig.4(a), we plot the phonon transmittance given by formula
(\ref{total}) for selected values of $\sigma^\prime$ along with the
numerical results for correlated disorder realizations.
Notice that in spite of the differences in the type of interfacial defect
the curves look very similar. A natural way to explain this result
is to view the interface for correlated disorder as a mosaic
of clusters of average diameter $\sim\xi$. Suppose that within
a particular cluster $i$ all sites are occupied by defects
with the same mass and characteristic
relative mass difference $\sigma^\prime_i$. As long as the phonon wavelength
is smaller than
the correlation length, i.e., $k \xi \gg 1$, the overall transmittance
can be considered as the average over all clusters with local transmittances
given by Eq.(\ref{total}). In Fig.4(b) such an average is plotted
for clusters with excess mass fully balanced by clusters with mass
deficit. The relative mass difference is taken from a Gaussian distribution
with width equal to the disorder strength $\sigma$. Clearly, this
procedure gives a better fit to the correlated disorder curve,
which suggests that the phonon transmittance across a correlated
disordered interface can be split to a statistically
averaged mean field contribution induced by mosaic grains and a fluctuating
component. The latter is controlled by the rapid residual variations
in the relative mass difference at the cluster boundaries as well as
the fluctuations in cluster sizes.

The qualitative agreement between correlated
disorder realizations and the mosaic picture prompts us to ask
whether there exists a simple model for the case of uncorrelated
disorder. To explore this possibility, we examine a two-colour disorder model
where linear interfacial defects are built up from
atoms with relative mass difference $\sigma^\prime$ which alternates sign
every $L/a$ sites. This model is motivated by the observation that
uncorrelated disorder contains all spatial harmonics with equal probability.
In what follows, we compare the exact transmittance for uncorelated disorder
with that of the two-colour model, averaged over
many values of $L$. In Fig.5, we plot the mean arithmetic value
of the phonon transmittances for $L=2,4,..,10a$
taken after averaging over 500 two-colour disorder
realizations given by a Gaussian distribution with zero mean
and $\sqrt{\langle {\sigma^\prime}^2 \rangle}  = \sigma$. The remarkable
agreement with the calculations for the uncorrelated disorder
over a wide range of frequencies suggests that the mean field
contribution for this case can be calculated by picking up few
elements from the basis set of the two-colour disorder model.

Finally, we compare the results of our calculation
with the diffuse mismatch model {\it DMM}
introduced  in \cite{Swartz} and extensively used in studies
of Kapitza conductance \cite{Pohl1,Swartz,Stoner}.
Application of the DMM at a boundary between two solids with identical
acoustic properties gives a $50\%$ phonon transmittance independent
of frequency. It can be easily demonstrated
that the Kapitza conductance $\sigma_K$ for
our model is larger than that predicted by the DMM.
It is also less than the radiation limit which in this case
would coincide with acoustic mismatch theory (total transmission).
Moreover, $\sigma_K$ is larger for uncorrelated disorder than for correlated.
Finally, one notes that the results of \cite{Maris} which are valid for phonon
transmission
through the ideal interface between two different fcc crystals a
transmittance which is frequency-independent over
the whole spectrum up to a cut-off frequency corresponding to maximum
phonon frequency in one of the crystals, whereas, in our
model the phonon transmittance is a strong function of frequency.

\section{Summary}
In summary, generic properties of elastic
phonon transport at a disordered interface
were studied by computing the exact scattering properties
of a single disordered atomic layer sandwiched
between identical waveguides.
The results show that phonon transmittance is a strong function of frequency
and the disorder correlation length. At frequencies lower than the
van Hove singularity the transmittance at a given frequency increases
as the correlation length decreases.
At low frequencies, this is reflected by different
power-laws for phonon conductance across correlated and uncorrelated disordered
interfaces which are in approximate agreement
with perturbation theory of an elastic continuum \cite{Alex}.
We also investigated simple mosaic and two-colour models of the interface
and showed that phonon transmittance through interfaces with
correlated and uncorrelated disorder can be understood in terms of
phonon conductances of structures with regular interfaces.

\newpage
\begin{figure}
\centerline{\psfig{figure=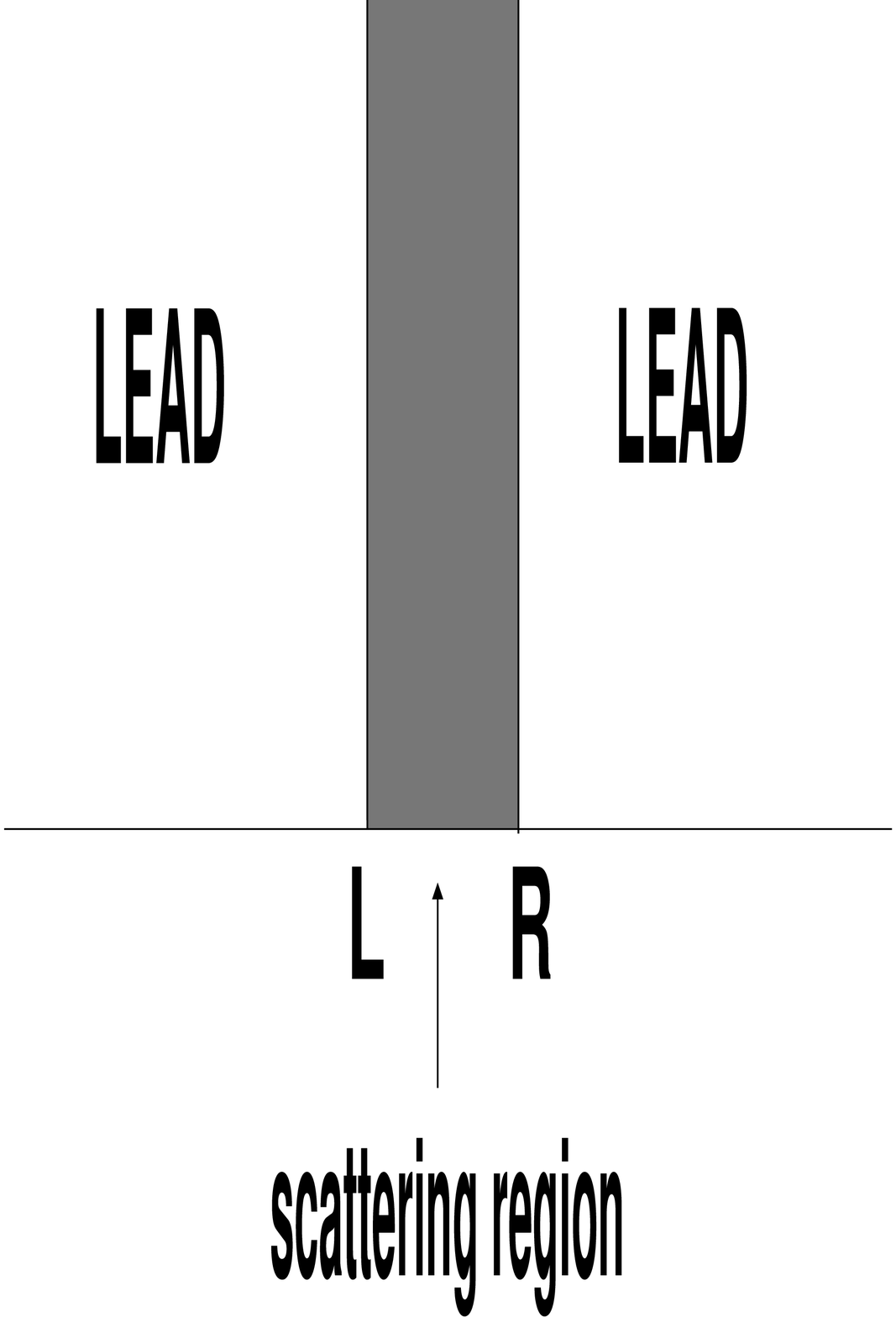,height=7.cm,width=10cm}}
\caption{A typical scattering geometry, where L(R) refer to sites
on the faces of left(right) lead.}
\label{struc}
\end{figure}

\newpage
\begin{figure}
\centerline{\psfig{figure=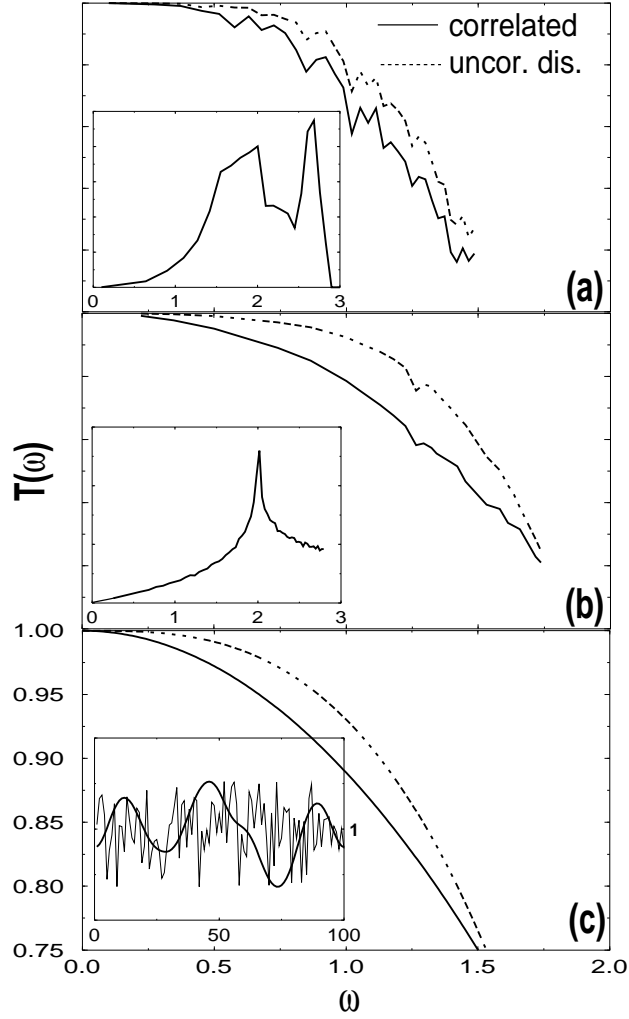,height=14.cm,width=10cm}}
\caption{Phonon transmittance for the (a)fcc model($\xi=5.65$),
(b)square scalar model($\xi =20$), (c)estimates based on perturbation theory.
In the (a) and (b) insets
the corresponding normalised DOS versus $\omega$ is plotted. Frequency is
measured
in units of $\sqrt{\kappa/m}$. Inset (c) shows a single realization
of $m_j$ versus j for a line of correlated and uncorrelated
masses($\sigma=0.3$ in all plots).}
\label{resul1}
\end{figure}

\newpage
\begin{figure}
\centerline{\psfig{figure=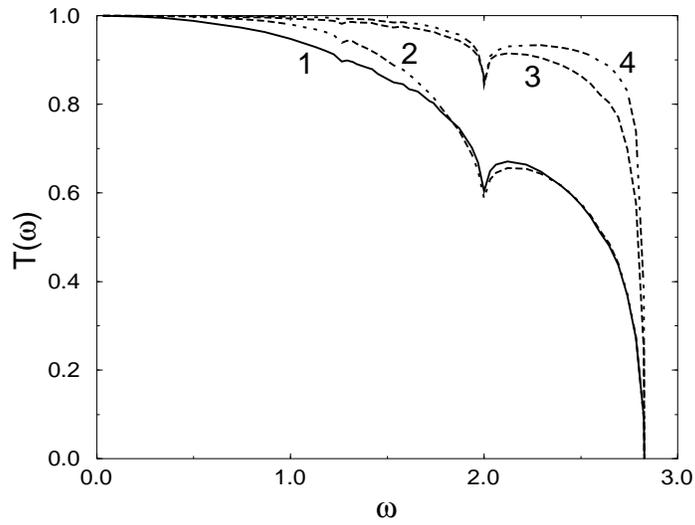,height=14.cm,width=10cm}}
\caption{Phonon transmittance for $\sigma=0.3$ and $0.1$. Curves (1), (2)
correspond to correlated ($\xi = 15$) and uncorrelated disorder for $\sigma=0.3$,
whereas curves (3), (4) are for  $\sigma=0.1$.}
\label{resul2}
\end{figure}

\newpage
\begin{figure}
\centerline{\psfig{figure=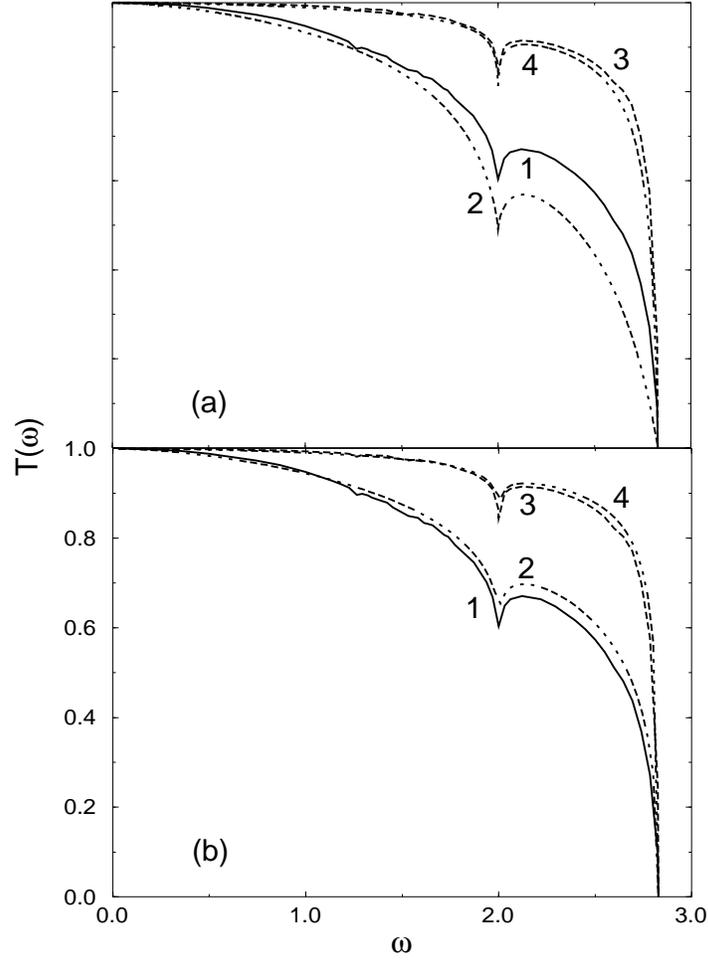,height=14.cm,width=10cm}}
\caption{Phonon transmittance for correlated disorder and
regular models as explained in the text.
In both (a) and (b) curves (1) and (3) correspond to correlated
disorder for $\sigma=0.3$ and $0.1$ ($\xi = 15$).
In (a), curves (2) and (4) are for the regular interlayer
with $\sigma^\prime=0.3$ and $0.1$. The results of the
mosaic model are shown in (2b) and (4b).}
\label{resul3}
\end{figure}

\newpage
\begin{figure}
\centerline{\psfig{figure=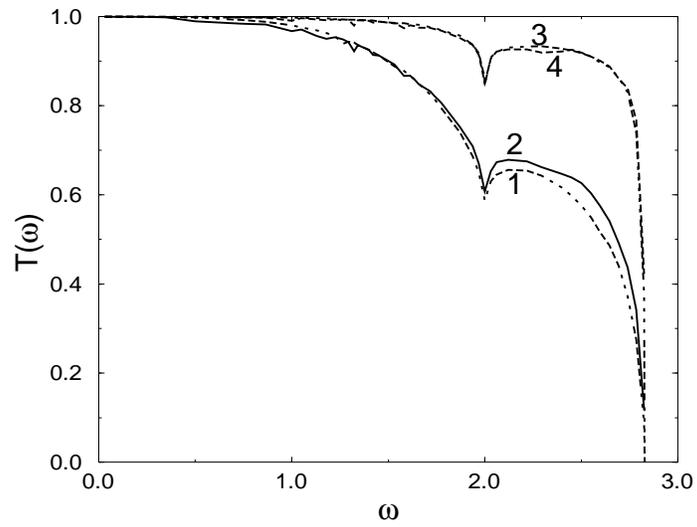,height=14.cm,width=10cm}},
\caption{Phonon transmittance for uncorrelated disorder,
curves (1) and (3) for  $\sigma=0.3$ and $0.1$,
and the two-colour disorder model.}
\label{resul4}
\end{figure}

\end{document}